\documentclass[onecolumn,showpacs,prc,superscriptaddress,amsmath,amssymb,preprintnumbers]{revtex4-1}

\newcommand{ \be }{\begin{eqnarray}}
\newcommand{ \ee }{\end{eqnarray}}

\usepackage{dcolumn} 
\usepackage{bm} 
\usepackage{graphicx}
\usepackage{color}
\usepackage[pdftex,bookmarks]{hyperref}
\usepackage{lineno}
\usepackage{graphicx}

\usepackage{amsmath}

\usepackage{amsfonts}

\usepackage{amssymb}

\usepackage{mathtools}

\definecolor{dgreen}{cmyk}{1.,0.,1.,0.4} 
\definecolor{orange}{cmyk}{0.,0.353,1.,0.} 

\begin{document}


%
\title{New estimator for symmetry plane correlations in anisotropic flow analyses}
\author{Ante Bilandzic} 
\affiliation{Physik Department, Technische Universit\"{a}t M\"{u}nchen, Munich, Germany}
\affiliation{Excellence Cluster Universe, Technische Universit\"{a}t M\"{u}nchen, Munich, Germany}
\author{Marcel Lesch} 
\affiliation{Physik Department, Technische Universit\"{a}t M\"{u}nchen, Munich, Germany}
\author{Seyed Farid Taghavi} 
\affiliation{Physik Department, Technische Universit\"{a}t M\"{u}nchen, Munich, Germany}
\date{\today}

\begin{abstract}
Correlations of symmetry planes are important observables used to quantify anisotropic flow phenomenon and constrain independently the properties of strongly interacting nuclear matter produced in the collisions of heavy ions at the highest energies. In this paper, we point out current problems of measuring correlations between symmetry planes and elaborate on why the available analysis techniques have a large systematic bias. To overcome this problem, we introduce a new approach to approximate multi-harmonic flow fluctuations via a two-dimensional Gaussian distribution. Employing this approximation, we introduce a new estimator, dubbed Gaussian Estimator (GE), to extract pure correlation between symmetry planes. We validate GE by using the realistic event-generator iEBE-VISHNU and demonstrate that it outperforms all existing estimators. Based on event-shape engineering, we propose an experimental procedure to improve GE accuracy even further.
\end{abstract}


\maketitle


\section{Introduction}
\label{s:Introduction}

The past years have witnessed the advent of large statistics heavy-ion datasets at RHIC and LHC facilities, comprising events with very large multiplicities. It is therefore becoming feasible to study the details of strongly interacting nuclear matter produced in heavy-ion collisions with unprecedented precision by employing multiparticle correlation techniques. When two heavy ions collide at ultrarelativistic energies a very rich and non-trivial sequence of stages emerges in the evolution of the produced fireball. Since each of these stages typically involves different underlying physics, ideally they would be described separately in theoretical models and probed one at a time in an experiment. To date, however, most of the analyzed heavy-ion observables are final-state observables in the momentum space, which pick up cumulatively the contributions from all stages in the heavy-ion evolution, starting all the way down from the details of the initial collision geometry. To leading order, these stages can be divided into the following categories: initial conditions, deconfined Quark-Gluon Plasma (QGP) stage, hadronization, chemical freeze-out, rescatterings, kinematic freeze-out, and finally free streaming. An important program in the field is the development of new observables which would be sensitive only to one particular stage in the heavy-ion evolution~\cite{Gale:2013da,Heinz:2013th,Braun-Munzinger:2015hba}.

For an idealized description of the heavy-ion collision geometry the initial volume containing interacting nucleons is ellipsoidal in non-central collisions. In such a case, anisotropic flow develops the shapes in the final-state momentum distribution which can be captured solely with the even Fourier amplitudes $v_{2n}$ and only one symmetry plane $\Psi_{\rm RP}$ (the reaction plane, spanned by the impact parameter vector and the beam axis)~\cite{Ollitrault:1992bk,Kolb:2003zi,Voloshin:2008dg}. However, in a more realistic description of the collision geometry, the initial energy density profiles fluctuate both in magnitude and in shape from one heavy-ion collision to another. Such initial-state fluctuations are also transferred into the final-state momentum fluctuations via anisotropic pressure gradients which develop in the fireball. Therefore, the full Fourier series expansion needs to be employed to quantify the anisotropies in the azimuthal distribution of emitted particles in the plane transverse to the beam axis:
\begin{equation}
f(\varphi) = \frac{1}{2\pi}\left[1+2\sum_{n=1}^\infty v_n\cos[n(\varphi-\Psi_n)]\right]\,,
\label{eq:Fourier_vn_psin}
\end{equation}
where $v_n$'s are anisotropic flow amplitudes, and $\Psi_n$'s the corresponding symmetry planes~\cite{Voloshin:1994mz}. In the past, anisotropic flow studies have been focused mostly on the flow amplitudes $v_n$. These results helped a great deal in establishing the perfect fluid paradigm about QGP properties~\cite{Heinz:2005zg,Akiba:2015jwa}.

From the above Fourier series expansion it can be seen immediately that, due to event-by-event flow fluctuations, $v_n$'s and $\Psi_n$'s are independent and equally important degrees of freedom to quantify anisotropic flow phenomenon, and therefore both sets of observables need to be studied and measured. However, the nature of these observables is different, which triggers the development of separate analyses techniques for their measurements. Further independent information about QGP and the other stages in the heavy-ion evolution can be extracted from the novel studies of observables which are sensitive to the intercorrelations between different flow amplitudes or between different symmetry planes, or from observables which couple the intercorrelations between both degrees of freedom. Another open question is the connection between $v_n$'s and $\Psi_n$'s defined in the final-state momentum distribution, and their counterparts in the initial coordinate space where they quantify the anisotropies stemming from the fluctuations of the initial collision geometry. This is mostly studied by the theorists because it is not feasible to access the initial stages of collision in the experiment.

Before starting to discuss the physics of $v_n$ and $\Psi_n$ observables, we first summarize the most important formal mathematical properties, which are used later in the derivation of our main results (additional details can be found in Appendix~\ref{a:Basic-properties-of-symmetry-planes}). Solely from the definition of Fourier series one can prove that $v_{-n} = v_n$ and $\Psi_{-n} = \Psi_n$, therefore in this paper we use them interchangeably. By combining the Fourier decomposition in Eq.~(\ref{eq:Fourier_vn_psin}) and the orthogonality properties of trigonometric functions, one can show that $v_n$'s and $\Psi_n$'s are related via the following mathematical identity:
\begin{equation}
v_n = \left<\cos[n(\varphi-\Psi_n)]\right>\,,
\label{eq:vn}
\end{equation}
where the average goes over all azimuthal angles $\varphi$ reconstructed in an event. Despite its simplicity, Eq.~(\ref{eq:vn}) has little relevance in experimental high-energy physics, due to difficulties in measuring reliably symmetry planes $\Psi_n$ in each heavy-ion collision. Instead, flow amplitudes $v_n$ can be estimated even without knowing the symmetry planes $\Psi_n$ by utilizing two- and multiparticle azimuthal correlations~\cite{Wang:1991qh,Jiang:1992bw,Borghini:2001vi,Bilandzic:2010jr,Bilandzic:2013kga}. Ollitrault {\it et al} have derived in~\cite{Bhalerao:2011yg} the most general relation between flow degrees of freedom $v_n$ and $\Psi_n$ and multiparticle azimuthal correlators, which is valid for any number of azimuthal angles $\varphi_1, \varphi_2, \ldots, \varphi_k$, and for any choice of harmonics $n_1, n_2, \ldots, n_k$:
\begin{equation}
v_{n_1}^{a_1}\cdots v_{n_k}^{a_k}\,e^{i(a_1 n_1\Psi_{n_1}+\cdots+a_k n_k\Psi_{n_k})} = \left<e^{i(a_1 n_1\varphi_1+\cdots+a_k n_k\varphi_k)}\right> \,.
\label{eq:generalResult}
\end{equation}
The average on the RHS in the above expression goes over all distinct tuples of $k$ azimuthal angles in an event. When compared to the original result from~\cite{Bhalerao:2011yg}, we have used a slightly different notation in the above expression by introducing $a_i$ coefficients which are by definition positive integers. The precise meaning of $a_i$ coefficient is the following: $a_i$ is the number of appearances of harmonic $n_i$ associated with different azimuthal angles in the azimuthal correlator on the RHS of Eq.~(\ref{eq:generalResult}) (positive and negative harmonics are counted separately). The advantage of this more general notation is that harmonics $n_i$ in Eq.~(\ref{eq:generalResult}) are now all unique by definition. In addition, $n_i$ and $a_i$ naturally split off when associated with flow amplitudes on LHS of Eq.~(\ref{eq:generalResult}), which makes their physical interpretation straightforward. It is straightforward to choose harmonics $n_1, n_2, \ldots, n_k$ in this general result in order to cancel the contribution from symmetry planes $\Psi_n$ and estimate solely flow amplitudes $v_n$ (e.g. the choice $n_1=n, n_2=-n, n_k = 0$ for $k > 2$, $a_1=a_2=1$, yields the standard formula for 2-particle azimuthal correlation $\left<\cos [n(\varphi_1\!-\!\varphi_2)]\right> = v_n^2$). However, it is much more of a challenge to derive an analogous expression that would express multiparticle azimuthal correlators only in terms of symmetry planes, i.e. the expression from which the prefactors $v_{n_1}^{a_1}\cdots v_{n_k}^{a_k}$ in Eq.~(\ref{eq:generalResult}) would cancel out exactly. So far in the literature only approximate relations have been presented in this context, all of which have the inherent systematic biases straight from their definitions. Such relations are particularly unreliable when used in the analyses of heavy-ion datasets characterized by large and correlated flow fluctuations, which is the case typically encountered in practice. In this paper, and as our main result, we introduce a new set of observables based on multiparticle azimuthal correlators, which are more reliable estimators of symmetry planes than the ones used previously. Before the presentation of our new results, we clarify for completeness sake which categories of observables depending only on symmetry planes are meaningful to study in practice. 

The fundamental difference between $v_n$ and $\Psi_n$ flow degrees of freedom lies in the fact that only $v_n$'s are invariant with respect to the arbitrary rotations of laboratory coordinate system in which azimuthal angles $\varphi$ are measured. We also remark that due to periodicity the symmetry plane angle $\Psi_n$ is uniquely determined only in the range $0 \leq \Psi_n < 2\pi/n$~\cite{Poskanzer:1998yz}. On the other hand, starting from Eq.~(\ref{eq:Fourier_vn_psin}) and using the constraint $0 \leq f(\varphi) \leq 1$ (probabilistic interpretation) it follows that each flow amplitude must satisfy $|v_n|<0.5$. Therefore, in order to eliminate trivial periodicity of each symmetry plane, and to ensure invariance of our observables with respect to random event-by-event fluctuations of the impact parameter vector, we arrive at the conclusion that the fundamental non-trivial observables involving symmetry planes are the following correlators and constraints~\cite{Bhalerao:2011yg,Jia:2012ju,Aad:2014fla}:
\begin{equation}
\left<e^{i(a_1n_1\Psi_{n_1} + \cdots + a_kn_k\Psi_{n_k})}\right>,\quad\sum_i a_i n_i = 0\,.
\label{eq:symmetry-plane-observables}
\end{equation}
The meaning of $a_i$ and $n_i$ is clarified in the text following Eq.~(\ref{eq:generalResult}). In the rest of the paper, we call observables in Eq.~(\ref{eq:symmetry-plane-observables}) Symmetry Plane Correlations (SPC). They can be estimated precisely only in theoretical models in which it is possible to compute each symmetry plane $\Psi_n$ for each heavy-ion collision. The main purpose of this paper is to establish a reliable experimental way to estimate SPC  indirectly by using only the azimuthal angles of reconstructed particles, since only they can be measured reliably in  experiments.

We conclude the introduction with a brief review of both experimental and theoretical results on SPC obtained so far in anisotropic flow analyses. We do not consider the evaluation of resolution factors in the standard event plane method, when symmetry planes corresponding to two or three subevents are used to correct for the effects of finite multiplicities, since such symmetry planes correspond to the same harmonic, only estimated in different subevents~\cite{Poskanzer:1998yz}. Also, we do not consider the correlations between symmetry planes and the reaction plane, since the latter cannot be estimated reliably in an experiment.

The importance of SPC in anisotropic flow measurements has been fully acknowledged only in the LHC era, even though the first results were obtained already 20 years ago in E877 experiment~\cite{Barrette:1996rs}. The first results at RHIC for SPC involving two symmetry planes were published by PHENIX in~\cite{Afanasiev:2009wq,Adare:2011tg}, by using the standard event plane method with subevent technique~\cite{Poskanzer:1998yz}. Both NA49 and STAR have analyzed 3-particle azimuthal correlators in mixed harmonics, which by definition do have contributions from symmetry planes, but their contribution has been neglected in these early analyses~\cite{Alt:2003ab,Adams:2003zg}. The first SPC studies at LHC provided only the binary statements on whether certain symmetry planes are correlated or not, without providing quantitative details---in~\cite{ALICE:2011ab} ALICE, by using the carefully designed 5-particle azimuthal correlator (the technical details can be found in Appendix~H of~\cite{Bilandzic:2012wva}), has demonstrated that the fluctuations of symmetry planes $\Psi_2$ and $\Psi_3$ are independent in all considered centralities. Finally, the most thorough experimental analysis to date, which included also the first measurements of correlations among three symmetry planes, has been published by ATLAS in~\cite{Jia:2012sa,Mohapatra:2012jr,Aad:2014fla}, by using the analysis technique discussed in~\cite{Jia:2012ma,Jia:2012ju}.

Theoretical studies have investigated SPC separately in coordinate (typically by using the Monte Carlo Glauber model~\cite{Miller:2007ri} in combination with event plane method) and momentum space~\cite{Alver:2010gr,Staig:2010pn,Nagle:2010zk,Jia:2012ma,Jia:2012ju,Qin:2011uw,Qiu:2012uy,Aad:2014fla,Yan:2015fva}. In these studies the values of symmetry planes are typically the direct output of the model in each heavy-ion collision, and therefore they do not need to be estimated indirectly by utilizing the azimuthal angles of produced particles. A notable independent approach to SPC in terms of conditional probabilities has been established in~\cite{Teaney:2010vd}. Other types of studies involving symmetry planes which we do not discuss in our paper have been performed in~\cite{Petersen:2010cw,Qin:2010pf,Lacey:2010av,Qiu:2011iv,Heinz:2013bua}. Finally, for the previous attempts to use azimuthal correlators to estimate SPC indirectly, we refer the reader to~\cite{Bhalerao:2011yg,Bhalerao:2011ry,Bhalerao:2011bp,Luzum:2012da,Bhalerao:2013ina}.

This paper is organized as follows. After introduction, in Section~\ref{s:Key-Idea} we present the key idea behind the new observables for SPC estimation, and point out the inherent systematic biases that plagued the previous approaches. Section~\ref{s:Gaussian-Estimator} discusses the concrete realization of our new estimators, dubbed Gaussian Estimator (GE). In Section~\ref{s:Event-Shape-Engineering} we present the comparison with the theoretical models, both for new and old SPC estimators, and indicate in which regime our estimators outperform the existing ones. Finally, in Section \ref{s:Conclusions} we summarize our results and outline the next steps. The technical details are provided in Appendices.


\section{Key idea of the new estimator for symmetry plane correlations}
\label{s:Key-Idea}

In this section we summarize the systematic biases of previous analyses that used azimuthal correlators to estimate SPC and introduce our new approach which improves on those biases. SPC were estimated previously by using the scalar product (SP) method~\cite{Voloshin:2008dg,vanderKolk:2012oca} or event plane method~\cite{Poskanzer:1998yz}, both of which yield the theoretical results for SPC only in the absence of correlated fluctuations of different flow magnitudes. Our new method, which we illustrate in the next paragraphs and elaborate in detail in Section~\ref{s:Event-Shape-Engineering}, provides a further step forward in a sense that it yields the theoretical result for SPC also when such correlated fluctuations of flow magnitudes are present in the data. In fact, at RHIC and LHC energies correlations of event-by-event fluctuations of $v_2$ and $v_3$, and of $v_2$ and $v_4$, are large and if they are not taken into account and corrected for, the final results for SPC can exhibit large systematic biases, as we demonstrate in Monte Carlo studies presented in Section~\ref{s:Event-Shape-Engineering}.

As already indicated in the introduction, correlations between $k$ symmetry planes in unique harmonics $n_1,...,n_k$ (i.e. correlations between $\Psi_{n_1},...,\Psi_{n_k}$) can be investigated by the measurement of the correlator $\left<\cos \left(a_{1} n_1 \Psi_{n_{1}} +\cdots + a_{k} n_k \Psi_{n_{k}} \right)\right>$, where the coefficients $a_{i}$ have to be fixed in such a way, that this expression is invariant in respect to the randomness of the reaction-plane. In theory, such correlators can be built from an event-by-event ratio of  two multiparticle azimuthal correlators. As a concrete example, by using the analytic formula in Eq.~(\ref{eq:generalResult}), one can derive the following result:
\begin{equation}
\frac{\left<\cos(2\varphi_1\!+\!2\varphi_2\!-\!\varphi_3\!-\!\varphi_4\!-\!\varphi_5\!-\!\varphi_6)\right>}{\left<\cos(2\varphi_1\!-\!2\varphi_2\!+\!\varphi_3\!-\!\varphi_4\!+\!\varphi_5\!-\!\varphi_6)\right>}=\frac{v_2^2v_1^4\cos 4(\Psi_{2}\!-\!\Psi_1)}{v_2^2v_1^4} = \cos 4(\Psi_{2}\!-\!\Psi_1)\,.
\label{eq:app-appetizer}
\end{equation}
This idea, which works only for correlators involving 6 or more azimuthal angles, demonstrates that for general ratios of this kind, the numerator consists out of both, flow amplitudes and symmetry planes, while the denominator only out of the respective flow amplitudes but without any contribution of symmetry planes. The correlators in numerator and denominator were carefully chosen so that the final expression depends only on the symmetry planes, even in the case of large correlated fluctuations of flow amplitudes $v_1$ and $v_2$.

We now generalize this starting example, and write in the most general case:
\begin{equation}\label{EbE}
\langle \cos\left(a_{1} n_1 \Psi_{n_1}+\cdots+a_{k} n_k \Psi_{n_k}\right) \rangle_{\text{EbE}}= \left< \frac{ v_{n_1}^{a_{1}} \;\cdots\; v_{n_k}^{a_{k}} \cos\left(a_{1} n_1 \Psi_{n_1}+\cdots+a_{k} n_k \Psi_{n_k}\right) + \delta }{ v_{n_1}^{a_{1}} \cdots v_{n_k}^{a_{k}}  + \delta'} \right>.
\end{equation}
As such, this event-by-event ratio exhibits only the symmetry planes. This remains true by definition also if the event-by-event fluctuations of flow amplitudes are correlated, and it is precisely this point which is not satisfied for the currently used estimators. Since both the numerator and denominator in Eq.~\eqref{EbE} have to be estimated with different $k$-particle azimuthal correlators ($k\geq 6$), they will have different statistical errors, which we denote by $\delta$ and $\delta'$, respectively. Such a direct event-by-event approach is at the moment experimentally not feasible due to large statistical uncertainties which prevent such a per-event ratio.

We overcome the limitation of event-by-event estimator in Eq.~\eqref{EbE} by introducing a new approximate method to estimate the SPC, which we refer to as the ``Gaussian Estimator'' (GE), in the next section. By using the same notation we now clarify the currently existing approximation methods, hereby focusing on the SP method. 
The explicit form of SP estimation is given by~\cite{Bhalerao:2013ina}
\begin{equation}\label{Eq. SP}
\langle \cos\left(a_{1} n_1 \Psi_{n_1}+\cdots+a_{k} n_k \Psi_{n_k}\right) \rangle_{\text{SP}}=\frac{\langle v_{n_1}^{a_{1}} \;\cdots\; v_{n_k}^{a_{k}} \cos\left(a_{1} n_1 \Psi_{n_1}+\cdots+a_{k} n_k \Psi_{n_k}\right)\rangle}{\sqrt{\langle v_{n_1}^{2a_{1}} \rangle \cdots \langle v_{n_k}^{2a_{k}} \rangle}}.
\end{equation}
The powers $a_{i}$ are chosen in such a way such that the numerator and the denominator are valid multiparticle correlators. We see later, that within the GE approximation the same kind of powers $a_{i}$ appear, and we provide a set of constraints which $a_i$ must satisfy in Appendix~\ref{Appendix: Choice of Correlators}.
We will see further that in most cases, the SP method is not an accurate estimator of the true SPC, since the denominator in Eq.~\eqref{Eq. SP} cannot be written in the factorized form. This statement is supported by experimental evidence of large correlations between the flow amplitudes \cite{Aad:2015lwa,ALICE:2016kpq,Sirunyan:2017uyl,STAR:2018fpo}, which therefore leads in general to the non-negligible bias in the SP method. On the other hand, estimators for SPC from the event plane method are plagued by finite resolutions in estimating each symmetry plane directly event-by-event~\cite{Aad:2014fla}.


\section{Gaussian Estimator}
\label{s:Gaussian-Estimator}

To begin with the GE approximation, we introduce the following quantities:
\begin{equation}\label{polar}
\mathcal{R}=  v_{n_1}^{a_1}\;\cdots\; v_{n_k}^{a_k},\qquad \Theta= a_1n_1 \Psi_{n_1}+\cdots+ a_kn_k \Psi_{n_k}.
\end{equation}
Using the above definitions, one can further define $R_x$ and $R_y$ in the Cartesian coordinate system as 
\begin{equation}
\mathcal{R}_x=\mathcal{R}\cos\Theta,\qquad \mathcal{R}_y=\mathcal{R}\sin\Theta.
\end{equation} 
Due to the event-by-event flow fluctuations, the quantities $\mathcal{R}_x$ and  $\mathcal{R}_y$ fluctuate from one event to the other. We can study the moments (or equivalently cumulants) of these fluctuations which are, in fact, the moments of probability density function (p.d.f.) $P(\mathcal{R}_x,\mathcal{R}_y)$ or equivalently $P(\mathcal{R},\Theta)$. One notes that for positive integers $p$ and positive even $q$, the moment $\langle \mathcal{R}^p_x\,\mathcal{R}^q_y\rangle = \langle\mathcal{R}^{p+q}\cos^p\Theta\sin^q\Theta \rangle$ is non-vanishing where the angular bracket indicates the average over events. The moments with odd $q$ are zero because of the presence of sin term with odd power. A simple example of such quantities is the case $k=2$, $n_1=-n_2=n$, and $a_1=a_2=\ell$ leading to $\mathcal{R}=v_n^{2\ell}$ and $\Theta=0$. In such a case, the moments $\langle v_n^{2\ell}\rangle$ have been extensively studied over the past years.
In general, the non-vanishing $\langle \mathcal{R}^p_x\,\mathcal{R}^q_y\rangle$ can be expanded in the basis spanned by the moments
\begin{equation}
\langle  m \rangle_{(a_1,n_1),\ldots,(a_k, n_k)}=v_{n_1}^{a_1}\cdots v_{n_k}^{a_k}\;e^{i(a_1 n_1 \Psi_{n_1}+\cdots+a_k n_k \Psi_{n_k})} , \qquad m=  \sum a_i,
\end{equation}
and estimated experimentally by employing multiparticle correlation techniques~\cite{Jiang:1992bw,Borghini:2001vi,Bilandzic:2010jr,Bilandzic:2013kga} (the above notation was introduced in the paragraph below Eq.~\eqref{eq:generalResult}). In the present study, we specially focus on the following moments, 
\begin{equation}\label{experimental}
\langle \mathcal{R}_{x} \rangle= \text{Re}\, \langle \langle  m \rangle_{(a_1,n_1),\ldots,(a_k, n_k)} \rangle,\qquad 
\langle \mathcal{R}^2_{x} \rangle+\langle \mathcal{R}^2_{y} \rangle=\left|\langle \langle 2 m \rangle_{(2a_1,n_1),\ldots,(2a_k, n_k)} \rangle\right|.
\end{equation}

We now develop the procedure to estimate the SPC, which in the current notation amounts to estimating  $\langle \cos\Theta \rangle$.
If we had been able to measure $P(\mathcal{R}_x,\mathcal{R}_y)$, we could have computed the moment $\langle \cos\Theta \rangle$ immediately. Although  the moments $\langle \mathcal{R}^p_x\,\mathcal{R}^q_y\rangle$ are accessible by multiparticle correlation techniques, the convergence of the moment expansion to the true value of p.d.f. is not guaranteed \cite{Stoyanov:2006}. Due to the central limit theorem, however, we are still able to approximately estimate the distribution as a 2D normal distribution, 
\begin{equation}\label{normal}
\mathcal{N}(\mathcal{R}_x,\mathcal{R}_y)=\frac{1}{2\pi\sigma_x\sigma_y}\exp\left[-\frac{(\mathcal{R}_x-\mu_x)^2}{2\sigma_x^2}-\frac{\mathcal{R}_y^2}{2\sigma_y^2}\right],
\end{equation}
where $\mu_x=\langle \mathcal{R}_{x} \rangle$, $\sigma_{x}^2=\langle \mathcal{R}^2_{x} \rangle-\langle \mathcal{R}_{x} \rangle^2$, and $\sigma_{y}^2=\langle \mathcal{R}^2_{y} \rangle$. 
Since we are only interested in angular part of the normal distribution in Eq.~\eqref{normal}, we integrate out the radial part. After some algebra, we find,
\begin{equation}
\mathcal{N}_\theta(\Theta)=\int \mathcal{R}\, d\mathcal{R} \; \mathcal{N}(\mathcal{R},\Theta)=\frac{\sigma_x^3\sigma_y\,e^{-\mu_x^2/2\sigma_x^2}}{\pi \sigma_\theta^2 }\left[1+\frac{\sqrt{\pi}\mu_x\sigma_y\;e^{\mu_\theta^2}}{\sigma_\theta}\left[1+\text{erf}\left(\mu_\theta\right)\right]\right],
\end{equation}
where
\begin{equation}
\sigma_\theta(\Theta)=\sigma_x\sqrt{2\sigma_y^2\cos^2\Theta+2\sigma_x^2\sin^2\Theta},\qquad \mu_\theta(\Theta)=\frac{\mu_x\sigma_y\cos\Theta}{\sigma_\Theta}.
\end{equation}
As a result, one can straightforwardly compute the average $\langle \cos\Theta \rangle$ by computing the following integral,
\begin{equation}\label{generalInt}
\langle \cos\Theta \rangle_{\text{GE}} = \int d\Theta\; \mathcal{N}_\Theta(\Theta) \cos\Theta,
\end{equation}
which is the Gaussian Estimator for the true value of $\langle \cos\Theta \rangle$. To find an analytical result for our estimator, we still need some simplifications as we discuss in what follows.

The quantity $\sigma_\theta$ has no $\Theta$ dependence by considering $\sigma_x\sim\sigma_y\sim\sigma_r/\sqrt{2}$ where $\sigma_r=\sqrt{\sigma_{x}^2+\sigma_y^2}$. This leads to an analytical result for integral in Eq.~\eqref{generalInt} written in terms of two first modified Bessel functions. By expanding the result in term of $\mu_x/\sigma_r$ and keeping only the leading term, we obtain
\begin{equation} \label{GaussApprox}
\begin{split}
\langle \cos\Theta \rangle_{\text{GE}}\simeq\sqrt{\frac{\pi}{4}}\left(\frac{\mu_x}{\sigma_r}\right).
\end{split}
\end{equation}
Using Eq.~\eqref{experimental}, the above approximation can be written explicitly as follows,
\begin{equation} \label{GaussApproxExp}
\begin{split}
&	\langle \cos\left(a_{1} n_1 \Psi_{n_1}+\cdots+ a_{k} n_k \Psi_{n_k}\right) \rangle_{\text{GE}}\simeq \sqrt{\frac{\pi}{4}}\;\;\;\frac{\langle  v_{n_1}^{a_{1}} \;\cdots\; v_{n_k}^{a_{k}}\cos\left(a_{1} n_1 \Psi_{n_1}+\cdots+a_{k} n_k \Psi_{n_k}\right) \rangle}{\sqrt{\langle  v_{n_1}^{2a_{1}} \;\cdots\;v_{n_k}^{2a_{k}}  \rangle}},
\end{split}
\end{equation}
where for the denominator we have used the fact that $\sigma_r=\sqrt{\langle \mathcal{R}^2_{x} \rangle-\mu_x^2+\langle \mathcal{R}^2_{y} \rangle}\simeq \sqrt{\langle \mathcal{R}^2_{x} \rangle+\langle \mathcal{R}^2_{y} \rangle}$. The error we have made in the second equality is of the order of $(\mu_x/\sigma_r)^2$. After comparing the above formula with Eq.~\eqref{Eq. SP}, one finds that apart from a numerical factor $\sqrt{\pi/4}\simeq0.886 $, we have a joint moment of flow amplitudes in the denominator. After we have introduced the technical details of GE approximation for SPC, in the next section we validate it by using realistic Monte Carlo simulation. We demonstrate that although the above approximation works accurately for most cases, there is still room to improve the accuracy by employing the event shape engineering. 


\section{Validation of Gaussian Estimator and its further improvements}
\label{s:Event-Shape-Engineering}

A new estimator for SPC has been introduced in the previous section by assuming that the $(\mathcal{R}_x,\mathcal{R}_y)$ fluctuation is approximately described by a 2D normal distribution. The accuracy and applicability of the method depend on this assumption. To examine the estimator's accuracy and in order to study possible ways for its improvements, we employ the realistic Monte Carlo event generator iEBE-VISHNU \cite{Shen:2014vra}. We initiate the events at $\tau=0.6$~fm/$c$ by MC-Glauber model \cite{Miller:2007ri}  implemented in the iEBE-VISHNU. For the hydrodynamic evolution DNMR \cite{Denicol:2010xn, Denicol:2012cn} causal hydrodynamic is solved at fixed shear viscosity over entropy density $\eta/s=0.08$ and the Cooper-Frye freeze-out \cite{Cooper:1974mv} prescription has been implemented in the package for particleization stage. The evolution in hadronic stage is not considered in our simulation. For each centrality bin, 14k events of Pb-Pb collisions ($\sqrt{\text{s}_{NN}} = 2.76$ TeV) have been generated and flow magnitudes $v_n$ and symmetry planes $\Psi_n$ are computed in each event for $\pi^\pm$, $K^\pm$ and $p/\bar{p}$ in the final state. The SPC obtained from these directly computed event-by-event symmetry planes are referred to as true value of SPC in the comparisons which we present next.

\begin{figure}[h]
	\centering
	\includegraphics[clip, trim=0cm 0cm 0cm 0cm, width=0.8\textwidth]{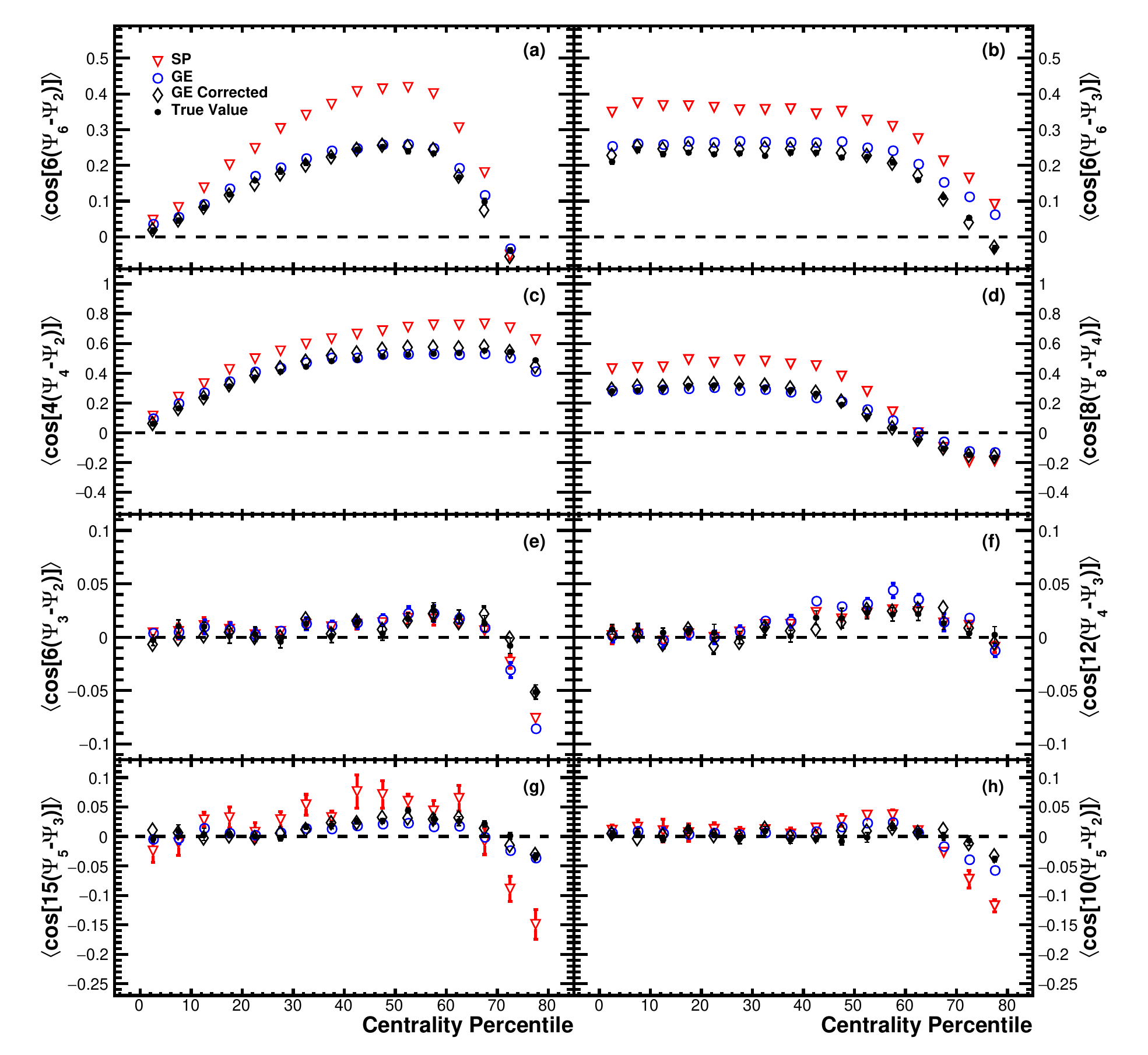}
	\caption{Comparison of GE and SP method to the true value of SPC between two symmetry planes in iEBE-VISHNU.}
	\label{Fig. 2SPC}
\end{figure}

\begin{figure}[h]
	\centering
	\includegraphics[clip, trim=0cm 0cm 0cm 0cm, width=0.8\textwidth]{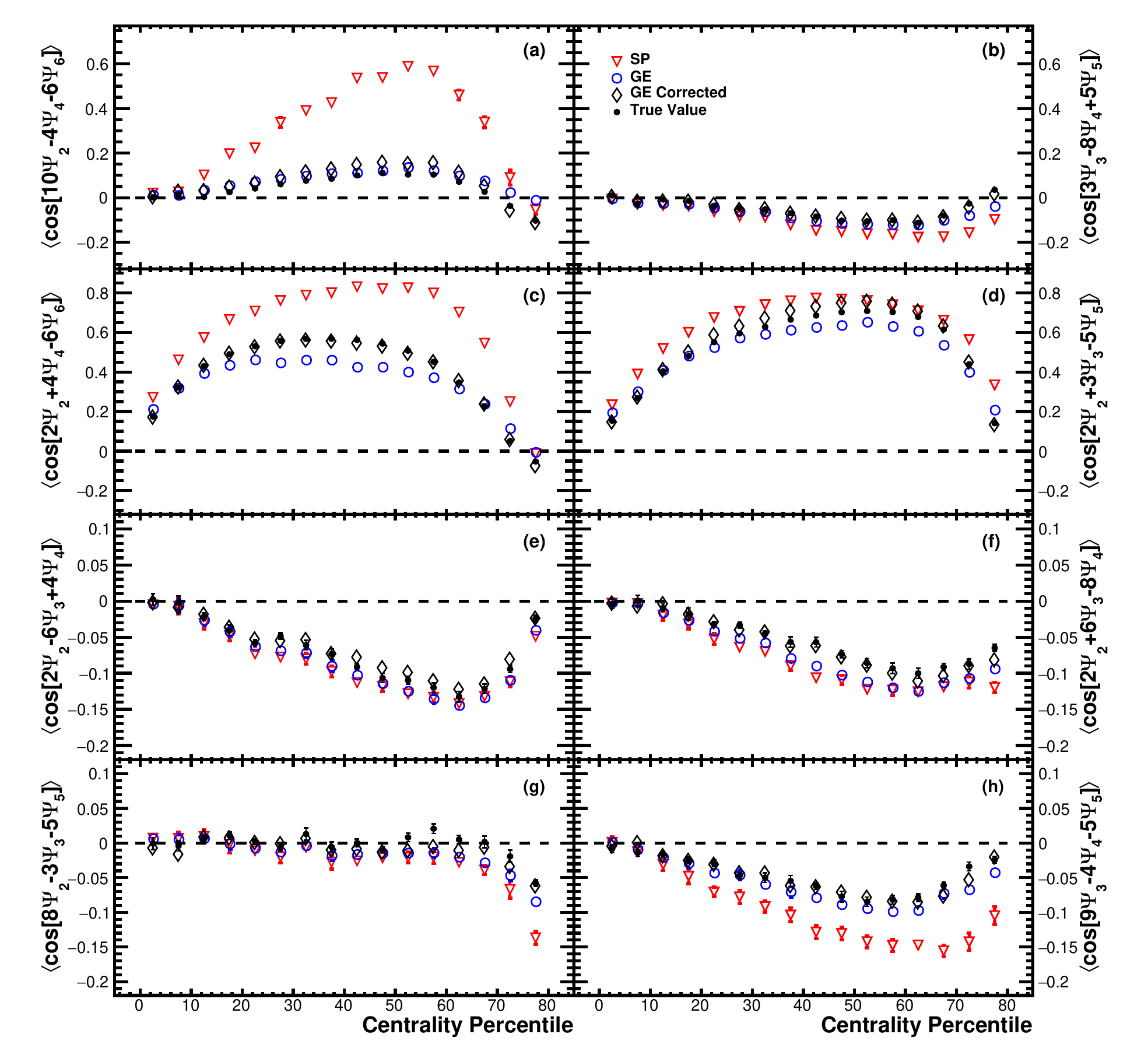}
	\caption{Comparison of GE and SP method to the true value of SPC between three symmetry planes iEBE-VISHNU.}
	\label{Fig. 3SPC}
\end{figure}

\begin{figure}[h]
	\centering
	\includegraphics[clip, trim=0.5cm 5cm 0.5cm 0cm, width=0.8\textwidth]{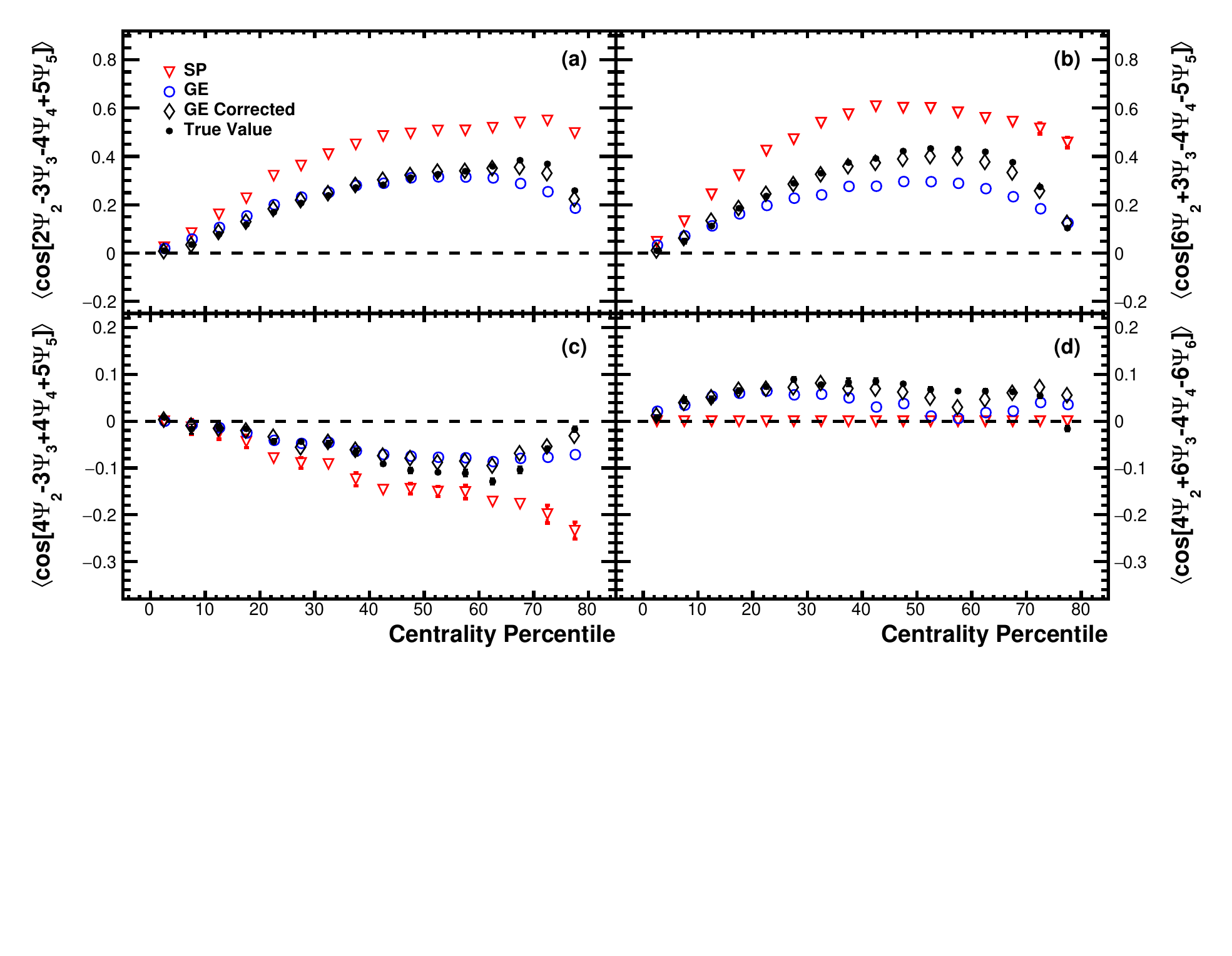}
	\caption{Comparison of GE and SP method to the true value of SPC between four symmetry planes iEBE-VISHNU.}
	\label{Fig. 4SPC}
\end{figure}

Our first study in Fig.~\ref{Fig. 2SPC} shows eight different choices for the correlation of two symmetry planes, and it demonstrates that the true value of SPC can be approximated much better with the GE approach, than with the SP estimator. Especially in cases, where the two symmetry planes are strongly correlated (Fig.~\ref{Fig. 2SPC}~(a)-(d)) due to their geometric correlations that pre-existed in the initial state (e.g. between $\Psi_2$ and $\Psi_4$), our new method reproduces the true value very well in all centrality classes of interest. This demonstrates clearly that the systematic bias caused by neglecting correlations between the flow amplitudes in the SP method is large, and therefore cannot be neglected. Only in a few cases it can be observed that the GE and SP yield comparable results (e.g. for SPC between $\Psi_4$ and $\Psi_3$). We will elaborate on this in more detail later and present a way to improve the GE method even further. The centrality dependence of each SPC in Fig.~\ref{Fig. 2SPC} presents strikingly different features, and therefore provides independent constraints for the system properties. Further, we present results for correlations between three  symmetry planes (Fig.~\ref{Fig. 3SPC}) as well as between four symmetry planes (Fig.~\ref{Fig. 4SPC}). It can be observed clearly that for each SPC the GE approach outperforms the SP estimator in most of considered centralities, while on the remaining few centralities the accuracy of the methods is comparable.

\begin{figure}[h]
	\centering
	\includegraphics[clip, trim=0cm 0cm 0cm 0cm, width=0.8\textwidth]{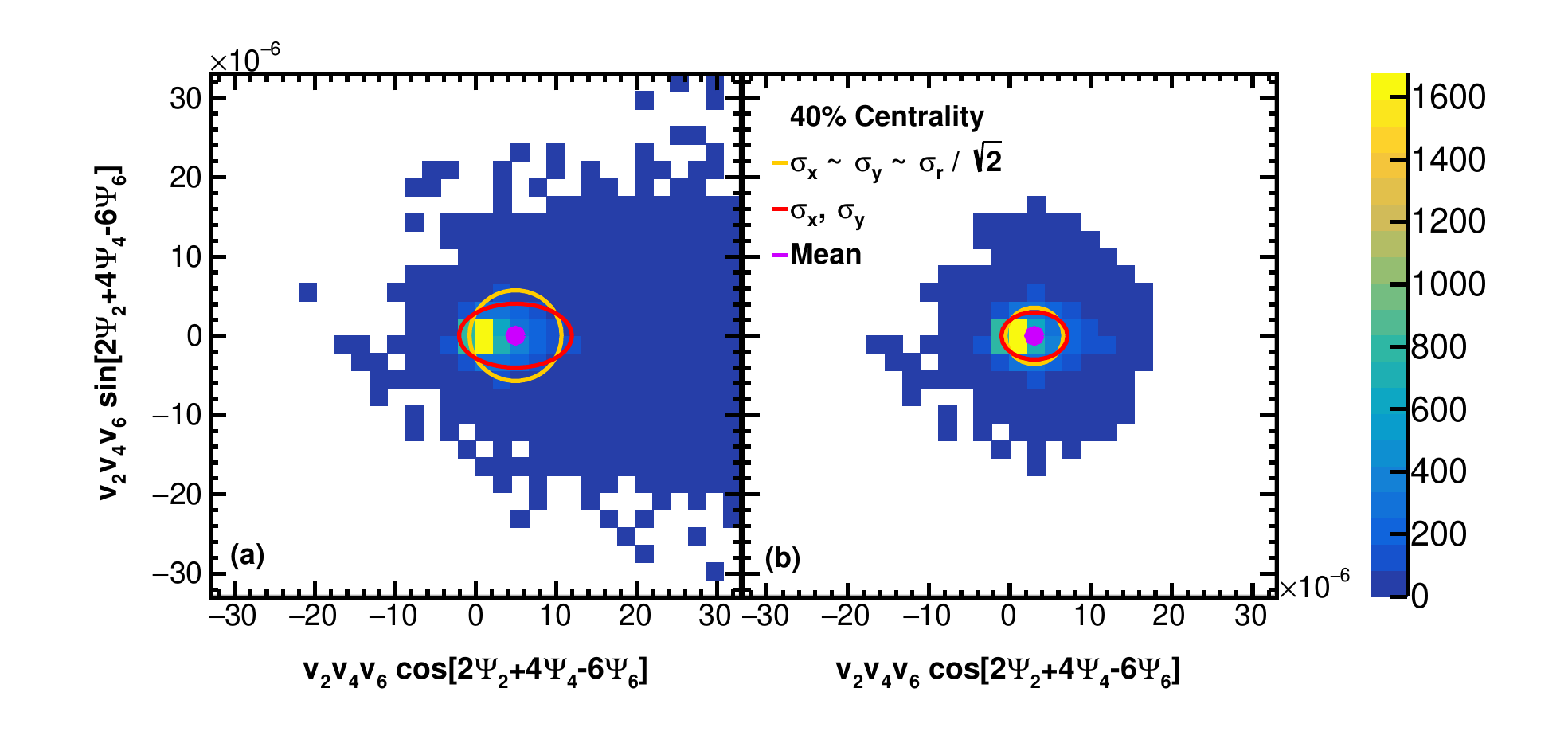}
	\caption{Distributions of $v_2v_4v_6 \cos\left[2\Psi_2+4\Psi_4-6\Psi_6\right]$ and $v_2v_4v_6 \cos\left[2\Psi_2+4\Psi_4-6\Psi_6\right]$ before (left) and after (right) correction by rejecting events bigger than $\mathcal{R}\leq \alpha\sigma_r$ ($\alpha=2$).}
	\label{Fig. GaussianDistHisto}
\end{figure}

Although the Gaussian Estimator in Eq.~\eqref{GaussApproxExp} works accurately for almost all cases, in contrast to the SP method which in most cases exhibits large systematic biases, there are still minor discrepancies between our estimator and the true value for few cases (see e.g. $\Theta=2\Psi_2+3\Psi_3-5\Psi_5$ in Fig.~\ref{Fig. 3SPC}~(d)). To investigate the reason more deeply, we focus on an extreme example: $\mathcal{R}=v_2v_4v_6$, $\Theta=2\Psi_2+4\Psi_4-6\Psi_6$ at 40\% centrality (see Fig. \ref{Fig. 3SPC}~(c)). In this case, there is a clear discrepancy between the true value and the Gaussian approximation Eq.~\eqref{GaussApproxExp}. In Fig.~\ref{Fig. GaussianDistHisto}, the iEBE-VISHNU outcome for $(\mathcal{R}_x,\mathcal{R}_y)$ fluctuations is shown. As it can be seen from the figure, there is a sharp peak at the center and a few events distributed around it. The tail is elongated in $x$-direction. Although there are much fewer events in the tail, it leads to inaccuracy in our Gaussian estimation. Specifically, the events are mostly concentrated symmetrically around the center while the long tail in the $x$-direction leads to a large difference between $\sigma_x$ and $\sigma_y$. Also, it shifts the $\mu_x$ to the right. The GE would work better if we could fit the Gaussian distribution around the peak and remove the outliers.

Since we have access to the value of flow magnitudes and symmetry planes in each event in the simulation, it is not a challenging task to remove the outliers. One can locate the peak in the histogram and fit a Gaussian distribution around it by ignoring events away from the peak with a certain criteria. Here, however, we try to introduce criteria that are model-independent and applicable also in experiments. We first compute $\sigma_x$ and $\sigma_y$ from all events. After that we divide the events into two classes: low $\mathcal{R}$ class with condition $\mathcal{R}\leq \alpha \sigma_r$ and the rest as high $\mathcal{R}$ class.  We have found that by ignoring the events at the tail of the distribution starting from twice the width $\sigma_r$ ($\alpha=2$), the GE is corrected very well as we will see shortly.
After event classification, we compute $\hat{\mu}_x$ and $\hat{\sigma}_r$ at low $\mathcal{R}$ class, and estimate $\langle \cos\Theta \rangle$ by using Eq.~\eqref{GaussApprox}. For the specific case shown in Fig.~\ref{Fig. GaussianDistHisto}~(a), the ratio $\sigma_x/\sigma_y$ computed from all events in the given centrality class is around 3 while if we compute the same ratio by using events in the low $\mathcal{R}$ class this ratio reduces to $1.7$. The corrected histogram with new $\hat{\mu}_x$, $\hat{\sigma}_x$, $\hat{\sigma}_y$, and $\hat{\sigma}_r$ is depicted in Fig.~\ref{Fig. GaussianDistHisto}~(b). The ``corrected'' Gaussian estimation $\langle \cos\Theta \rangle$ are shown in Figs.~\ref{Fig. 2SPC}~-~\ref{Fig. 4SPC} with open diamond markers indicating an improvement in most cases.
This classification is simple in the simulation while experimentally one needs to employ more sophisticated techniques such as event-shape engineering \cite{Schukraft:2012ah}. It is worth mentioning that the low $\mathcal{R}$ class contains 94\% of all events in our simulation. This means by removing 6\% of high $\mathcal{R}$ events the ratio  $\sigma_x/\sigma_y$ is reduced approximately by a factor of two. Experimentally, one is able to classify events into low and high $\mathcal{R}$ classes with $\sim 6$\% of events at high $\mathcal{R}$ class. The classification percentile can be optimized by comparing the low $\mathcal{R}$ class ratio $\sigma_x/\sigma_y$ with that obtained from all events in the given centrality class.


\section{CONCLUSIONS}
\label{s:Conclusions}

After introducing the new procedure to correct for the correlated flow fluctuations of different flow magnitudes, we have reduced significantly the systematic biases in the existing experimental techniques for symmetry plane correlations. This correction emerged from the modeling of experimentally accessible moments with a 2D Gaussian distribution. By using this new method, dubbed Gaussian Estimator, we have shown a significant improvement over existing SPC measurements in most cases of interest. We have demonstrated that in combination with event shape engineering, this new estimator can be optimized even further.

The precision measurements of SPC in the future have to acknowledge the remaining small intercorrelation between flow amplitudes and symmetry planes, which can still cause a small bias in all available approximation methods for SPC measurements.	


\acknowledgments{
	This project has received funding from the European Research Council (ERC) under the European Unions Horizon 2020 research and innovation programme (grant agreement No 759257).
}


\appendix


\section{Basic properties of symmetry planes}
\label{a:Basic-properties-of-symmetry-planes}

In this Appendix we outline in more detail the most important formal properties of symmetry planes. Besides the version of the Fourier series presented in the introduction in Eq.~(\ref{eq:Fourier_vn_psin}), the alternatively used form is:
\begin{equation}
f(\varphi)=\frac{1}{2\pi}\big[1+2\sum_{n=1}^\infty (c_n\cos n\varphi + s_n\sin n\varphi)\big]\,,
\label{eq:Fourier_cn_sn}
\end{equation}
with
\begin{eqnarray}
c_n &=& \int_0^{2\pi}f(\varphi)\cos(n\varphi) d\varphi\,,\label{c_n}\\
s_n &=& \int_0^{2\pi}f(\varphi)\sin(n\varphi) d\varphi\,,\label{s_n}
\end{eqnarray}
The Fourier series parametrizations in Eqs.~(\ref{eq:Fourier_vn_psin}) and (\ref{eq:Fourier_cn_sn}) are mathematically equivalent and can be interchanged by using the following relations:
\begin{eqnarray}
v_n&\equiv&\sqrt{c_n^2+s_n^2}\,,\label{eq:vn_definition}\\
\Psi_n&\equiv&(1/n)\arctan\frac{s_n}{c_n}.
\label{eq:Psin_definition}
\end{eqnarray}
The relation (\ref{eq:Psin_definition}) can be used as a definition of symmetry plane $\Psi_n$. We discuss next some physical properties of symmetry planes, and establish the connection between them and some commonly used observables in anisotropic flow analyses.

The symmetry plane $\Psi_n$ has an obvious geometrical interpretation when the anisotropic distribution can be parameterized only with one harmonic $n$, since then one can show immediately that 
\begin{equation}
f(\Psi_n+\varphi) = f(\Psi_n-\varphi)\,,
\end{equation}
i.e. a symmetry plane $\Psi_n$ is the plane for which it is equally probable for a particle to be emitted above and below it. From Eq.~(\ref{eq:Psin_definition}) one can see that symmetry planes are meaningful only when $c_n \neq 0$. If the flow amplitude $v_n$ is 0, or if the Fourier series permits only the sin term $s_n$, the corresponding symmetry plane $\Psi_n$ does not exist.  Similarly, odd symmetry planes $\Psi_{2n+1}$ do not exist in a system in which the exact symmetry is $f(\varphi) = f(\varphi+\pi)$, since that symmetry sets all odd $c_{2n+1}$ harmonics to zero.
Another symmetry of interest is $f(\varphi) = f(-\varphi)$ due to which $s_n = 0$ for all $n$, and therefore from Eq.~(\ref{eq:Psin_definition}) $\Psi_n = 0\ \forall n$, i.e. all symmetry planes are the same and equal to 0. Physically, this means that a heavy-ion collision was described in the laboratory frame with the coordinate system oriented such that the impact parameter vector is aligned with the $x$-axis. The next symmetry which is to leading order satisfied in non-central heavy-ion collisions is $f(\varphi) = f(\pi+\varphi)$, due to which $c_{2n+1}, s_{2n+1} = 0$ and therefore only the even symmetry planes $\Psi_{2n}$ are well-defined and non-trivial. In principle, one could also consider the symmetry $f(\varphi) = f(\pi-\varphi)$ in mid-central collisions, but we were not able to extract any new constraint on the symmetry planes, which was not already covered by the other symmetries. Finally, since we assign to $f(\varphi)$ the probabilistic interpretation (which implies that $f(\varphi)$ must be a positive definite function), we do not consider symmetries like $f(\varphi) = -f(-\varphi)$, which otherwise could lead to additional constraints. 

Another important physical interpretation of symmetry planes can be drawn from their relation with the $Q$-vector~\cite{Ollitrault:1992bk,Voloshin:1994mz,Barrette:1994xr}, which is one of the most important objects in flow analyses. For a set of $M$ azimuthal angles $\varphi_i$, the $Q$-vector in harmonic $n$ is defined as:
\begin{equation}
Q_n \equiv \sum_{j=i}^M e^{in\varphi_j}\equiv |Q_n|e^{in\Psi_n}\,.
\label{eq:Q-vector-definition}
\end{equation}
With such a definition, one can easily demonstrate that the angle of the $Q$-vector is exactly the same as symmetry planes $\Psi_n$ from Fourier series defined before in Eq.~(\ref{eq:Psin_definition}), since:
\begin{eqnarray}
(1/n)\arctan\frac{s_n}{c_n} &=&(1/n)\arctan\frac{\left<\sin n\varphi\right>}{\left<\cos n\varphi\right>}\nonumber\\
&=&(1/n)\arctan\frac{M{\rm Im} (Q_n)}{M {\rm Re} (Q_n)}\nonumber\\
\nonumber\\
&=&(1/n)\arctan\frac{|Q_n| \sin n\Psi_n}{|Q_n| \cos n\Psi_n}\nonumber\\
&=&(1/n)\arctan \tan n\Psi_n\nonumber\\
&=&(1/n) n\Psi_n\nonumber\\
&=&\Psi_n\,.
\label{eq:symmetryPlanesAndQvectors}
\end{eqnarray}
This relation is utilized in the standard event plane method, where symmetry planes $\Psi_n$ are estimated directly from $Q$-vectors in each event~\cite{Poskanzer:1998yz}.


\section{Choice of Correlators \label{Appendix: Choice of Correlators}}
In this Section we will start from the most general form of multiparticle correlators with non-unique harmonics from which on we will find constraints such that these correlators are applicable for our GE method (Eq.~\eqref{GaussApproxExp}). We will see that from there on, constraints for the $a_i$ will emerge naturally. \\
Consider two general multi-particle correlators $\left< k\right>_{n_1, n_2, ... , n_k}$ ($k$-particle correlator with set of non-unique harmonics $\{n_1$, $n_2$, ... $n_k\}$) and $\left< l\right>_{p_1, p_2, ... , p_l}$ ($l$-particle correlator with set of non-unique harmonics $\{p_1$, $p_2$, ... , $p_l\}$). Focusing on the general form of the GE approximation (Eq.~\eqref{GaussApproxExp}), their ratio can in general be written as
\begin{equation}
\frac{\left<\left< k\right>_{n_1, n_2, \cdots , n_k}\right>}{\left<\left< l\right>_{p_1, p_2, \cdots , p_l}\right>} \propto \frac{\left<v_{n_1}\cdots v_{n_k}e^{i(n_1\Psi_{n_{1}}+\cdots+n_k\Psi_{n_k})}\right>}{\sqrt{\left<v_{p_1}\cdots v_{p_l}e^{i(p_1\Psi_{p_{1}}+\cdots+p_l\Psi_{p_l})}\right>}} \,.
\end{equation}
From this general ansatz the following constraints to achieve the desired SPC emerge
\begin{align}
&\sum_{j=1}^{k} n_j  = 0 \label{Constraint1} \\
&\sum_{j=1}^{l} p_j = 0  \label{Constraint2} \\
&\sum_{j=1}^{k} n_j \cdot \Psi_{n_j} \neq 0 \label{Constraint3} \\
&\sum_{j=1}^{l} p_j \cdot \Psi_{p_j} = 0 \label{Constraint4} \\
&\prod_{i=1}^{k}v_{n_i}^2  = \prod_{i=1}^{l}v_{p_i} \,. \label{EQHarmonics}
\end{align}
Constraints~\eqref{Constraint1} and \eqref{Constraint2}  satisfy the isotropy condition which has to hold true for any non-trivial multi-particle correlator. Constraints~\eqref{Constraint3}  and \eqref{Constraint4}  lead to a non-vanishing contribution of symmetry planes in the numerator while the denominator does not depend on symmetry planes explicitly.  Constraint~\eqref{EQHarmonics}  ensures that the flow amplitudes in numerator and denominator cancel each other exactly. Further, from Constrain~\eqref{EQHarmonics} it follows that $l=2k$. Therefore, while measuring a $k$-particle correlator in the numerator one has to measure a $2k$-particle correlator in the denominator, when using GE approximation. To obtain the SPC  one has to explicitly choose sets of correlators  $\{n_1$, $n_2$, ... $n_k\}$ and $\{p_1$, $p_2$, ... , $p_l\}$ which satisfy constraints  Eqs.~\eqref{Constraint1}~-~\eqref{EQHarmonics}. We elaborate on this now explicitly for the SPC between two symmetry planes $\Psi_m$ and $\Psi_n$, and demonstrate how the coefficients $a_i$ used in the main part, see e.g. Eq.~(\ref{eq:generalResult}), emerge naturally and which constraints $a_i$ have to fulfil themselves. This formalism can be generalized for correlations between any amount of symmetry planes. 
\subsection{Correlators between two symmetry planes \label{Appendix: 2-SPC}}
Focussing now on the SPC between two symmetry planes $\Psi_m$ and $\Psi_n$ and given the  constraints Eq.~\eqref{Constraint3} to Eq.~\eqref{EQHarmonics}, the general sets of correlators in harmonics $m$ and $n$ (where $m\neq n$) are schematically
\begin{eqnarray}
&\left\{ \underbrace{m}_{a_m \,\, times},\cdots,m, \underbrace{-n}_{a_n \,\, times},\cdots,-n \right\} \quad& ({\mathrm{numerator}}) \label{eq: 2SPC Numerator Set} \\
&\left\{ \underbrace{m,-m}_{2 a_m \,\, times},\cdots,m,-m, \underbrace{n,-n,}_{2 a_n \,\, times},\cdots,n,-n \right\} \quad& ({\mathrm{denominator}}) \label{eq: 2SPC Denominator Set}
\end{eqnarray}
where $a_m, a_n \in \mathbb{N}$. Given the constraints Eq.~\eqref{Constraint1} and Eq.~\eqref{Constraint2} the following constraints for $a_m$ and $a_n$ are valid
\begin{equation}
\sum_{j=1}^{ a_m} m + \sum_{k=1}^{  a_n} (-n) =  a_m  m -  a_n  n = 0 \implies \frac{a_m}{n} = \frac{a_n}{m}\,, \label{Eq A Constraint1}
\end{equation}
\begin{equation}
\sum_{j=1}^{2  a_m} (-1)^j \cdot m + \sum_{k=1}^{2 a_n} (-1)^k \cdot n = 0 \implies 2 a_m \wedge 2 a_n \, \text{even} \label{Eq A Constraint2}
\end{equation}
where $\wedge$ is the logical AND.  This way, the constraints from Eq.~\eqref{Constraint3} and Eq.~\eqref{Constraint4} are satisfied as well. We see that Constraint~\eqref{Eq A Constraint2} will hold true for any $a_m, \; a_n$. Therefore, as the concrete example one can choose $a_m$ and $a_n$ as 
\begin{align}
a_m = \frac{l_{mn}}{m}  \\
a_n  =  \frac{l_{mn}}{n}
\end{align}
where $l_{mn}$ denotes the least common multiple between $m$ and $n$. The order of particle correlator in the numerator is given as 
\begin{equation}
l_{mn} \left(\frac{1}{m} + \frac{1}{n}\right)
\end{equation}
and for the numerator twice the size respectively. This method of using the least common multiple presents the lowest order of valid multiparticle correlators for the SPC between two symmetry planes. Any other method exhibits higher order of correlators. Given by this, the GE approach reads
\begin{equation}
\left<\cos \left[l_{mn} \left( \Psi_m - \Psi_n \right) \right]\right>_{\text{GE}} \propto \frac{\left<v_m^{a_m}v_n^{a_n} \cos \left[l_{mn} \left( \Psi_m - \Psi_n \right) \right]\right>}{\sqrt{\left<v_m^{2a_m}v_n^{2a_n}\right>}} \,.
\end{equation}
Although the method of the least common multiple exhibits the lowest possible order for a SPC with two planes, any multiple $k \in \mathbb{N}$ of this method represents a valid correlator as well. We can therefore always expand the set of correlators by changing $a_m \rightarrow ka_m$ and $a_n \rightarrow ka_n$ and there find in general
\begin{equation}
\left<\cos \left[kl_{mn} \left( \Psi_m - \Psi_n \right) \right]\right>_{\text{GE}} \propto \frac{\left<v_m^{ka_m}v_n^{ka_n} \cos \left[kl_{mn} \left( \Psi_m - \Psi_n \right) \right]\right>}{\sqrt{\left<v_m^{2ka_m}v_n^{2ka_n}\right>}} \,.
\end{equation}
\subsection{Correlators between three symmetry planes}
A general choice for the set of correlators for three unique harmonics $m$, $n$ and $p$ are schematically
\begin{eqnarray}
&\left\{ \underbrace{m}_{a_m \,\, times},\cdots,m, \underbrace{-n}_{a_n \,\, times},\cdots,-n, \underbrace{-p}_{a_p \,\, times},\cdots,-p \right\} \quad& ({\mathrm{numerator}}) \label{eq: 3SPC Numerator Set} \\
&\left\{ \underbrace{m,-m}_{2 a_m \,\, times},\cdots,m,-m, \underbrace{n,-n,}_{2 a_n \,\, times},\cdots,n,-n, \underbrace{p,-p,}_{2 a_p \,\, times},\cdots,p,-p  \right\} \quad& ({\mathrm{denominator}}) \label{eq: 3SPC Denominator Set}
\end{eqnarray}
Following the general constraints presented above we find the following constraints on $a_m$, $a_n$ and $a_p$ 
\begin{equation}
\sum_{j=1}^{ a_m} m + \sum_{k=1}^{  a_n} (-n)  + \sum_{l=1}^{  a_p} (-p) =  a_m  m -  a_n  n - a_p  p = 0  \label{Eq A 3SPC Constraint1}
\end{equation}

\begin{equation}
\sum_{j=1}^{2  a_m} (-1)^j \cdot m + \sum_{k=1}^{2 a_n} (-1)^k \cdot n +  \sum_{l=1}^{2 a_p} (-1)^l \cdot p = 0 \implies 2 a_m \wedge 2a_n \wedge 2a_p \, \text{even} \label{Eq A 3SPC Constraint2}
\end{equation}
Again the latter constraint is fulfilled trivially. In general these kind of correlators will be of high order, therefore limiting experimental feasibility. We cannot reduce the problem of a 3-SPC into one single closed formula as it has been the case for two planes, as now more combinatorial possibilities exist. As a trivial example, in cases that $m = n + p$  we can set trivially $a_m = a_n = a_p = 1$ \\
\begin{equation}
\left<\cos \left[ m\Psi_m - n\Psi_n - p\Psi_p  \right]\right>_{\text{GE}} \propto \frac{\left<v_mv_n v_p \cos \left[ m\Psi_m - n\Psi_n - p\Psi_p  \right]\right>}{\sqrt{\left<v_m^{2}v_n^{2}v_p^{2}\right>}} \,.
\end{equation}



\end{document}